\documentclass[10pt,twosides,a4paper]{book}
\usepackage[utf8]{inputenc}
\usepackage{bbm}
\usepackage{bm}
\usepackage{physics} 
\usepackage{amssymb}
\usepackage[pdftex]{graphicx}
\usepackage{latexsym}
\usepackage[small]{caption2}
\usepackage{amsmath}
\usepackage[english]{babel}
\usepackage{babel}
\usepackage{amsmath,amsthm}
\usepackage{rotating}
\usepackage{multirow}
\usepackage{graphicx}
\usepackage{tocloft}
\usepackage{bbm}
\usepackage{dsfont}
\usepackage{amsfonts}
\usepackage{accents}
\usepackage{pdfpages}
\usepackage[thinc]{esdiff}
\usepackage{hyperref}
\usepackage[numbers,sort&compress,square]{natbib}
\usepackage{natbib}
\usepackage{pdfsync}
\usepackage{fancyhdr}
\usepackage{scalerel,stackengine}
\usepackage{fancyhdr}
\usepackage{booktabs}
\renewcommand{\cftchappresnum}{CHAPTER }
\newlength{\mylen}
\fancypagestyle{plain}{} 

\input{tcilatex} 

\begin{document} 
\setcounter{chapter}{3}

\chapter[Alexandre P. Costa and Alexandre Dodonov \newline
{\em Quantum Rabi
oscillations in the semiclassical limit: backreaction on the cavity field
and entanglement}]{Quantum Rabi
oscillations in the semiclassical limit: backreaction on the cavity field
and entanglement}
\label{chapter4}

\markboth{Quantum Rabi oscillations in the semiclassical limit}{A. P. Costa and A.
Dodonov}

{\large \textbf{Alexandre P. Costa$^1$ and Alexandre Dodonov$^{1,2,a}$}}

\vspace{3mm}


\noindent {$^1$ Institute of Physics, University
of Brasilia, 70910-900, Brasilia, DF, Brazil}

\noindent {$^2$ International Center of Physics, Institute of Physics, University
of Brasilia, 70910-900, Brasilia, DF, Brazil}

\noindent {\texttt{$^{a}$adodonov@unb.br}}


\section{Abstract}

The goal of this chapter is to compare the predictions of the semiclassical
Rabi model (SRM), which describes the interaction between a two-level system
(qubit) and a classical monochromatic wave, and the quantum Rabi model
(QRM), under the assumption that the cavity field is initiated in a coherent
state with a large average number of photons, ranging from $5.000$ to $%
40.000 $. First, we show that for a strong atom--field coupling, when the
duration of the $\pi $-pulse (the time interval required to completely
excite or deexcite the qubit in the resonant regime) is below $100\omega
^{-1}$, the behaviour of the atomic excitation probability deviates
significantly from the textbook sinusoidal formula derived for the SRM under
the rotating-wave approximation, and we present simple analytical and
semi-analytical methods to describe more accurately the dynamics. Then we
show that the QRM reproduces the qubit's dynamics predicted by the SRM only
for initial times, since in the QRM the qubit excitation probability
exhibits a collapse behaviour even in the lossless scenario; we also notice
that the qualitative behaviour of such collapses is different from the ones
occurring in the dissipative SRM. In the rest of this work we study
numerically the backreaction of the qubit on the cavity field and the
resulting atom--field entanglement, which are disregarded in the SRM. It is
shown that the atom--field entanglement increases over time and a maximally
entangled state is attained for large times. Moreover, we illustrate how the
Rabi oscillations continuously modify the quantum state of the cavity field,
which becomes increasingly different from the original coherent state as the
time increases.

\section{Introduction}

Rabi oscillations or Rabi flopping of a two-level system (that we also call
qubit, or atom, for short) constitute an invaluable tool in the area of
Quantum Information and Quantum Computing, as they allow for the
implementation of single-qubit quantum gates, readout of the qubit's state,
measurement of the coherence properties of qubits, etc~\cite{dodon.mi,dodon.mat,dodon.calib}.
The Rabi oscillations, first studied by I. I. Rabi in the context of a spin
in a classical rotating magnetic field~\cite{dodon.rabi1,dodon.rabi2}, correspond to the
periodic oscillation of the atomic population inversion due to a drive by a
resonant or near-resonant classical field, and take place in a plethora of
physical systems, such as nuclear magnetic resonance, cold atoms, impurity
states in insulators, quantum dots, superconducting circuits, Rydberg atoms,
cavity polaritons, etc.~\cite{dodon.merlin}. It is probably one of the first
examples of the light--matter interaction one learns when studying Quantum
Optics or Atomic Physics, since a simple analytical solution can be obtained
after invoking the Rotating Wave Approximation (RWA)~\cite{dodon.boyd,dodon.scully,dodon.shore}%
. The mathematical description of the interaction between a quantum
two-level system and a periodically time-varying classical field has become
known as the semiclassical Rabi model (SRM), and many analytic approaches
were developed to determine its spectral and dynamical properties beyond the
RWA~\cite{dodon.graham,dodon.munz,dodon.Liu,dodon.Lu,dodon.ashhab} or in the presence of external modulations
\cite{dodon.castanos,dodon.ad3,dodon.ad7,dodon.marinho1,dodon.marinho3}.

When the driving field is treated quantum mechanically, the mathematical
description is given by the quantum Rabi model (QRM)~\cite{dodon.rev,dodon.solano,dodon.braak,dodon.rson}%
, and the transition from the quantum to the semiclassical regime as the
field intensity increases has been recently investigated in~\cite{dodon.shus1,dodon.shus2}. Since we know that the electromagnetic field is quantized, it
is obvious that the SRM is an idealization of the actual interaction between
the qubit and a coherent field state with large amplitude; hence the qubit
must become entangled with the field during the evolution (at the very
least, during some time intervals), and also modify the field state. Here we
study, first, the degree of entanglement between the qubit and the cavity
field in the semiclassical limit, in order to assess the intrinsic
limitations for the purity of the qubit (due to the quantized nature of the
electromagnetic field) during the operation of quantum gates based on the
Rabi flopping. Second, we analyse the modifications of the cavity field
state due to the Rabi oscillations, discussing whether the semiclassical
assumption that the field remains forever in the initial coherent state is
justifiable. Our findings, based on numeric simulations involving relatively
large average photon numbers and long time intervals, can be used to
estimate the situations in which the SRM is sufficiently accurate to
describe the qubit behaviour.

\section{Semiclassical Rabi model}

Consider a two-level system with the ground and excited states $|g\rangle $
and $|e\rangle $, respectively, and the energy difference $\hbar \Omega $,
where $\hbar $ is the Planck constant and $\Omega $ is the atomic transition
frequency. The interaction of the atom with a classical monochromatic
electromagnetic field of frequency $\omega $ is described by the celebrated
semiclassical Rabi Hamiltonian~\cite{dodon.merlin,dodon.boyd,dodon.scully,dodon.shore}
\begin{equation}
\hat{H}/\hbar =\frac{\Omega }{2}\hat{\sigma}_{z}+G\left( \hat{\sigma}_{+}+%
\hat{\sigma}_{-}\right) \cos \left( \omega t\right) \,,  \label{dodon.H}
\end{equation}%
where the semiclassical light--matter coupling parameter $G$ is known as the
Rabi frequency, and the atomic operators are $\hat{\sigma}_{z}=|e\rangle
\langle e|-|g\rangle \langle g|$, $\hat{\sigma}_{+}=|e\rangle \langle g|$
and $\hat{\sigma}_{-}=\hat{\sigma}_{+}^{\dagger }=|g\rangle \langle e|$. To
derive the qubit dynamics one needs to solve the Schr\"{o}dinger equation%
\begin{equation}
i\frac{\partial |\psi \rangle }{\partial t}=\frac{\hat{H}}{\hbar }|\psi
\rangle \,.
\end{equation}%
Making the unitary transformation, $|\psi (t)\rangle =e^{-i\omega t\hat{%
\sigma}_{z}/2}|\psi _{1}(t)\rangle \,$, the new wavefunction $|\psi
_{1}\rangle $ obeys the Schr\"{o}dinger equation with the Hamiltonian%
\begin{equation}
\hat{H}^{(1)}=\hat{H}_{r}+\hat{H}_{cr}\,,
\end{equation}%
where%
\begin{eqnarray}
\hat{H}_{r}/\hbar &=&-\frac{\Delta }{2}\hat{\sigma}_{z}+\frac{G}{2}\left(
\hat{\sigma}_{+}+\hat{\sigma}_{-}\right) \\
\hat{H}_{cr}/\hbar &=&\delta \frac{G}{2}\left( e^{2it\omega }\hat{\sigma}%
_{+}+e^{-2it\omega }\hat{\sigma}_{-}\right) \,.
\end{eqnarray}%
$\hat{H}_{r}$ is the time-independent \textquotedblleft
rotating\textquotedblright\ Hamiltonian, and $\hat{H}_{cr}$ is the
time-dependent \textquotedblleft counter-rotating\textquotedblright\
Hamiltonian due to the counter-rotating terms. We also defined the detuning $%
\Delta =\omega -\Omega $ and the dichotomous parameter $\delta =\left\{
0,1\right\} $: for $\delta =1$ we have the complete case, while $\delta =0$
implies the RWA.

The eigenvalues of $\hat{H}_{r}$ are easily found as $E_{\pm }=\pm \hbar R/2$%
, where $R=\sqrt{G^{2}+\Delta ^{2}}$. The corresponding eigenstates are%
\begin{equation}
|\phi _{\pm }\rangle =\sqrt{\frac{1}{RR_{\pm }}}\left( R_{\pm }|g\rangle \pm
\frac{G}{2}|e\rangle \right) \,,  \label{dodon.fip}
\end{equation}%
where we defined $R_{\pm }=\left( R\pm \Delta \right) /2$. In particular, in
the \emph{resonant regime}, $\Delta =0$, we have $R=G$, $R_{\pm }=G/2$ and $%
|\phi _{\pm }\rangle =\left( |g\rangle \pm |e\rangle \right) /\sqrt{2}$.
From Eq.~(\ref{dodon.fip}) one can easily calculate all the matrix elements $%
\langle \phi _{l}|\hat{\sigma}_{i}|\phi _{k}\rangle $, where $l,k=\pm $ and $%
\hat{\sigma}_{i}$ stand for any qubit operator.

The solution of the Schr\"{o}dinger equation for the Hamiltonian $\hat{H}%
^{(1)}$ can be found by expanding the system state in terms of the
eigenstates of $\hat{H}_{r}$ as~\cite{dodon.marinho2}%
\begin{equation}
|\psi _{1}\rangle =e^{-iE_{-}t}\left[ e^{-iRt}A_{+}(t)|\phi _{+}\rangle
+A_{-}(t)|\phi _{-}\rangle \right] \,\,,  \label{dodon.a0}
\end{equation}%
where $A_{\pm }(t)$ are the probability amplitudes, and for the initial
state $|\psi (0)\rangle =c_{g}|g\rangle +c_{e}|e\rangle $ one has
\begin{equation}
A_{\pm }\left( 0\right) =\sqrt{\frac{1}{RR_{\pm }}}\left( R_{\pm }c_{g}\pm
\frac{G}{2}c_{e}\right) \,.  \label{dodon.ica}
\end{equation}%
From Eq.~(\ref{dodon.a0}), the atomic excitation probability reads%
\begin{equation}
P_{e}=\left\vert \langle e|\psi (t)\rangle \right\vert ^{2}=\left( \frac{G}{%
2R}\right) ^{2}\left\vert \sqrt{\frac{R}{R_{+}}}A_{+}(t)e^{-iRt}-\sqrt{\frac{%
R}{R_{-}}}A_{-}(t)\right\vert ^{2}\,.
\end{equation}%
Analogously, all the other atomic properties can be determined from the
knowledge of $|\psi _{1}\rangle $.

Under the RWA, $\delta =0$, the probability amplitudes $A_{\pm }$ become
time-independent, and for the exact resonance, $\Delta =0$, one obtains the
textbook expression~\cite{dodon.boyd,dodon.scully}%
\begin{equation}
P_{e}=\sin ^{2}\frac{Gt}{2}+\left\vert c_{e}\right\vert ^{2}\cos Gt-\func{Im}%
\left( c_{e}c_{g}^{\ast }\right) \sin Gt\,\,.  \label{dodon.rwa}
\end{equation}%
In particular, for the \textquotedblleft $\pi $-pulse\textquotedblright ,
corresponding to the time interval $T_{\pi }=\pi /G$, the complete
transition $|g\rangle \leftrightarrow |e\rangle $ is accomplished.

Without the RWA, setting $\delta =1$, one has to solve the differential equations
\begin{equation}
i\dot{A}_{+}=\frac{G^{2}}{2R}\cos \left( 2\omega t\right) A_{+}+e^{itR}\frac{%
G}{2R}\left( R_{-}e^{2i\omega t}-R_{+}e^{-2i\omega t}\right) A_{-}
\end{equation}%
\begin{equation}
i\dot{A}_{-}=-\frac{G^{2}}{2R}\cos \left( 2\omega t\right) A_{-}+e^{-itR}%
\frac{G}{2R}\left( R_{-}e^{-2i\omega t}-R_{+}e^{2i\omega t}\right) A_{+}\,.
\end{equation}%
Defining new time-dependent probability amplitudes $a_{\pm }(t)$ via the
transformation%
\begin{equation}
A_{\pm }(t)=e^{\mp i\Upsilon \sin \left( 2\omega t\right) /2}a_{\pm }(t)\,~,
\label{dodon.a2}
\end{equation}%
where we defined the dimensionless parameter%
\begin{equation}
\Upsilon \equiv \frac{G^{2}}{2\omega R}\,\,,
\end{equation}%
we obtain the exact differential equations%
\begin{equation}
\dot{a}_{+}=-iQ_{t}a_{-}\quad ,\quad \dot{a}_{-}=-iQ_{t}^{\ast }a_{+}
\label{dodon.apm}
\end{equation}%
with%
\begin{equation}
Q_{t}=\frac{G}{2R}e^{i\left[ \Upsilon \sin \left( 2\omega t\right) +Rt\right]
}\left( R_{-}e^{2i\omega t}-R_{+}e^{-2i\omega t}\right) \,.  \label{dodon.qtr}
\end{equation}

In this work we focus on the resonant regime, $\Delta =0$, for which $%
\Upsilon =G/2\omega $ and
\begin{equation}
Q_{t}=\frac{G}{4}e^{i\Upsilon \cos \left( 2\omega t-\pi /2\right)
}e^{iRt}\left( e^{2i\omega t}-e^{-2i\omega t}\right) \,.
\end{equation}%
For the solution of Eqs.~(\ref{dodon.apm}) near the multi-photon resonances, $%
\Omega \approx \left( 2n+1\right) \omega $, see~\cite{dodon.ad7,dodon.marinho2}. In our
present case of one-photon resonance, from the Jacobi-Anger expansion we
write%
\begin{equation}
e^{i\Upsilon \cos \theta }=J_{0}\left( \Upsilon \right) +\sum_{n=1}^{\infty
}i^{n}J_{n}\left( \Upsilon \right) \left( e^{i\theta n}+e^{-i\theta
n}\right) ~,  \label{dodon.ja}
\end{equation}%
where
\begin{equation}
J_{n}\left( \Upsilon \right) =\sum_{k=0}^{\infty }\frac{\left( -1\right) ^{k}%
}{k!\left( n+k\right) !}\left( \frac{\Upsilon }{2}\right) ^{n+2k}
\end{equation}%
is the Bessel function of the first kind, with the known property $%
J_{n}\left( -x\right) =\left( -1\right) ^{n}J_{n}\left( x\right) $. Assuming
the realistic condition $\Upsilon \ll 1$ and keeping only the three initial
terms in the expansion~(\ref{dodon.ja}), we obtain%
\begin{equation}
Q_{t}\approx i\frac{G}{2}\left[ \left( J_{0}-J_{2}\right) \sin 2\omega
t+iJ_{1}\right] e^{iGt}\,,  \label{dodon.qt}
\end{equation}%
where we used the short-hand notation $J_{k}\equiv J_{k}\left( \Upsilon
\right) $. By solving numerically Eqs.~(\ref{dodon.apm}) with $Q_{t}$ given by Eq.
(\ref{dodon.qt}) one finds the atomic excitation probability%
\begin{equation}
P_{e}=\frac{1}{2}\left\vert e^{-i\left[ \Upsilon \sin \left( 2\omega
t\right) +Gt\right] }a_{+}-a_{-}\right\vert ^{2}\,.  \label{dodon.sa}
\end{equation}

We call this an approximate \emph{semi-analytic solution}, since $a_{\pm }$
still has to be determined by solving somehow Eqs.~(\ref{dodon.apm}) together
with $Q_{t}$ given by Eq.~(\ref{dodon.qt}). A simple closed approximate analytic
solution, better than the RWA solution, Eq.~(\ref{dodon.rwa}), can be obtained by
neglecting the rapidly oscillating term proportional to $\sin 2\omega t$ in
Eq.~(\ref{dodon.qt}). The resulting differential equations%
\begin{equation}
\dot{a}_{\pm }=i\frac{G}{2}J_{1}e^{\pm iGt}a_{\mp }~
\end{equation}%
can be easily cast as second-order differential equations%
\begin{equation}
\ddot{a}_{\pm }\mp iG\dot{a}_{\pm }+\left( \frac{GJ_{1}}{2}\right)
^{2}a_{\pm }=0\,.
\end{equation}%
The solutions read%
\begin{equation}
a_{+}\left( t\right) =e^{iGt/2}\left[ b_{1}e^{iGst/2}+b_{2}e^{-iGst/2}\right]
\label{dodon.w1}
\end{equation}%
\begin{equation}
a_{-}\left( t\right) =\frac{e^{-iGt/2}}{J_{1}}\left[ b_{1}\left( 1+s\right)
e^{iGst/2}+b_{2}\left( 1-s\right) e^{-iGst/2}\right] \,,  \label{dodon.w2}
\end{equation}%
where $s=\sqrt{1+J_{1}^{2}}$. The coefficients $b_{1}$ and $b_{2}$ are
determined from the initial condition~(\ref{dodon.ica}) as%
\begin{eqnarray}
b_{1} &=&\frac{\left( s-1+J_{1}\right) c_{g}+\left( s-1-J_{1}\right) c_{e}}{2%
\sqrt{2}s} \\
b_{2} &=&\frac{\left( s+1-J_{1}\right) c_{g}+\left( s+1+J_{1}\right) c_{e}}{2%
\sqrt{2}s}.
\end{eqnarray}%
For the lack of a better name, we call this solution an \emph{%
\textquotedblleft intermediate\textquotedblright\ solution}, since it
partially takes into account the counter-rotating terms.

\begin{figure}[tbh]
\begin{center}
\includegraphics[width=0.99\textwidth]{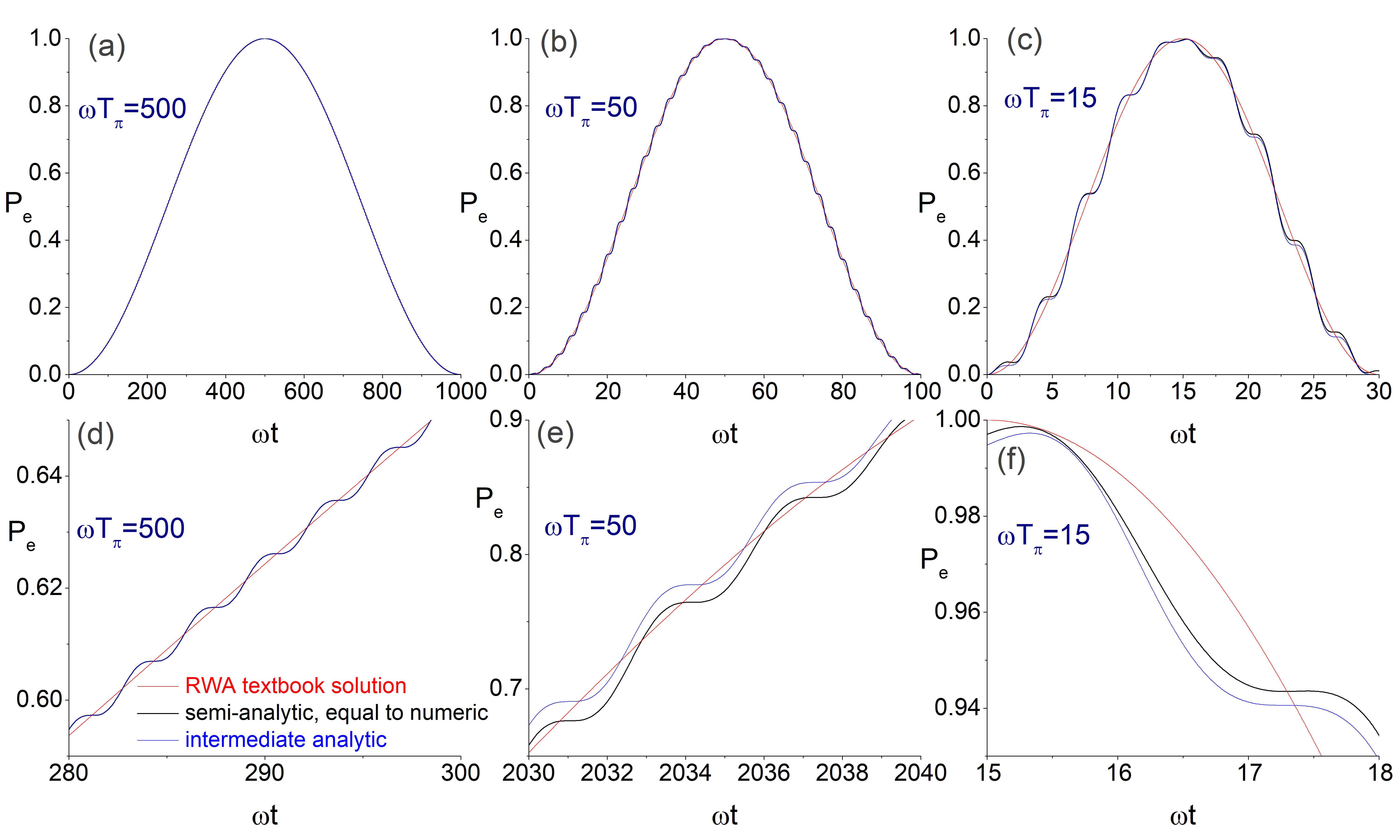} {}
\end{center}
\caption{Atomic excitation probability as function of the dimensionless time
for the lossless SRM. The Rabi frequencies, $G=\protect\pi /T_{\protect\pi }$%
, are parametrized in terms of $T_{\protect\pi }$ as: a) $T_{\protect\pi %
}=500\protect\omega ^{-1}$. b) $T_{\protect\pi }=50\protect\omega ^{-1}$. c)
$T_{\protect\pi }=15\protect\omega ^{-1}$. Red lines depict the RWA
solution. Black lines depict the semi-analytic solution, which is
indistinguishable from the exact numeric solution for the chosen parameters.
Blue lines depict the intermediate analytic solution. d-f) Zooms of the
panels (a-c) illustrating the discrepancies between different solutions.}
\label{dodon.fig1}
\end{figure}

To test the validity of the approximate methods, we solved numerically Eqs.~(
\ref{dodon.apm}) --~(\ref{dodon.qtr}), which describe exactly the dynamics of the SRM,
and compared the results to the predictions of the RWA, intermediate, and
semi-analytical solutions. We considered the resonant regime, $\Delta =0$,
and the initial ground state, $|\psi (0)\rangle =|g\rangle $. Figure \ref%
{dodon.fig1} shows the behaviour of the atomic excitation probability $P_{e}$ as a function of time for three different coupling strengths, parametrized in
terms of the duration of the $\pi $-pulse: $\omega T_{\pi }=500$ (panel~\ref{dodon.fig1}a, with the zoom in the panel~\ref{dodon.fig1}d), $\omega T_{\pi }=50$
(panel~\ref{dodon.fig1}b, with zoom in panel~\ref{dodon.fig1}e) and $\omega T_{\pi }=15$
(panel~\ref{dodon.fig1}c, with zoom in panel~\ref{dodon.fig1}f); the corresponding Rabi
frequencies read $G/\omega =\pi /(\omega T_{\pi })$. The red lines show the
standard RWA solution, Eq.~(\ref{dodon.rwa}), while the black lines depict the
semi-analytical solution, which for the considered parameters is
indistinguishable from the exact numeric solution. The blue lines depict the
intermediate solution, which is quite precise for relatively small coupling
strengths, $\omega T_{\pi }\gtrsim 50$, and initial times, $\omega t\lesssim
10^{3}$, but loses accuracy as the time increases. This can be clearly
seen in the panels~\ref{dodon.fig1}d~--~\ref{dodon.fig1}f, which show the discrepancies
between different approximate methods. On the other hand, for very short
pulses, $\omega T_{\pi }\lesssim 15$, the intermediate solution is
inaccurate even for the initial times, $t\sim T_{\pi }$, while the
semi-analytic solution holds perfectly. These results by no means imply that
one cannot use the resonant light--matter interaction to control the qubit
state. Instead, they show that to accurately control the atomic state one
cannot rely on the standard RWA expression for \textquotedblleft
short\textquotedblright\ pulses with $\omega T_{\pi }\lesssim 500$, and a
more accurate approach is necessary (e.g., the semi-analytic or the exact
numeric solutions).

However, for an accurate preparation and control of the qubit state, as
required in many protocols in the areas of Quantum Information and Quantum
Computing, one also needs to know the degree of qubit's purity. One obvious
reason for non-unitary purity is the dissipative effects, which arise from
the interaction of the qubit with its surroundings, so the qubit dynamics
must be described by some master equation for the density operator $\hat{\rho%
}$. A fairly accurate approach consists in employing the \textquotedblleft
standard\textquotedblright\ Markovian master equation of Quantum Optics~\cite%
{dodon.ad8,dodon.vogel}%
\begin{equation}
\frac{\partial \hat{\rho}}{\partial t}=-i\left[ \hat{H}/\hbar ,\hat{\rho}%
\right] +\frac{\gamma }{2}(n_{th}+1)\mathcal{D}[\hat{\sigma}_{-}]\hat{\rho}+%
\frac{\gamma }{2}n_{th}\mathcal{D}[\hat{\sigma}_{+}]\hat{\rho}+\frac{\gamma
_{\phi }}{2}\mathcal{D}[\hat{\sigma}_{z}]\hat{\rho}\,,  \label{dodon.eme}
\end{equation}%
where $\gamma $ is the damping rate, $\gamma _{\phi }$ is the pure dephasing
rate, $\mathcal{D}[\hat{\sigma}_{k}]\hat{\rho}\equiv 2\hat{\sigma}_{k}\hat{%
\rho}\hat{\sigma}_{k}^{\dagger }-\hat{\sigma}_{k}^{\dagger }\hat{\sigma}_{k}%
\hat{\rho}-\hat{\rho}\hat{\sigma}_{k}^{\dagger }\hat{\sigma}_{k}$ is the
Lindblad superoperator and $n_{th}=\left[ e^{\hbar \Omega /k_{B}T}-1\right]
^{-1}$ is the average thermal photon number for the reservoir's temperature $%
T$ ($k_{B}$ is the Boltzmann constant). This equation will be used in the
next section with $n_{th}=0.05$ to calculate numerically the purity of the
qubit, as well as the actual dynamics in the dissipative case (for an
approximate solution of this master equation near multi-photon resonances
see~\cite{dodon.marinho1,dodon.marinho2,dodon.marinho3}). Another reason for non-unitary
purity is the inevitable entanglement between the atom and the cavity field
due to their interaction -- a phenomenon overlooked in the semiclassical
model, that requires the quantized description of the cavity field. So in
the next section we shall study numerically the QRM to find out how the
qubit and the field get entangled throughout the evolution and how this fact
drastically affects the purity (and the overall behavior) of the system.

\section{Quantum Rabi model}

The quantum Rabi model~\cite{dodon.rev,dodon.solano,dodon.braak} is described by the
Hamiltonian%
\begin{equation}
\hat{H}_{QRM}/\hbar =\frac{\Omega }{2}\hat{\sigma}_{z}+\omega \hat{n}%
+g\left( \hat{a}+\hat{a}^{\dagger }\right) \left( \hat{\sigma}_{+}+\hat{%
\sigma}_{-}\right) \,,  \label{dodon.QRM}
\end{equation}%
where $\omega $ is the resonant frequency of the cavity, $\hat{a}$ and $\hat{%
a}^{\dagger }$ are the annihilation and creation operators of the
electromagnetic field, $\hat{n}=\hat{a}^{\dagger }\hat{a}$ is the photon
number operator and $g$ is the one-photon light-matter coupling constant.
Although the QRM is exactly solvable~\cite{dodon.braak}, the analytic solution is
cumbersome, and in many cases the counter-rotating terms $\left( \hat{a}\hat{%
\sigma}_{-}+\hat{a}^{\dagger }\hat{\sigma}_{+}\right) $ are neglected to
obtain the easily solvable Jaynes-Cummings Hamiltonian~\cite{dodon.scully,dodon.orszag}.
But here we solve numerically the complete Hamiltonian~(\ref{dodon.QRM}).

Let us briefly see how the quantum Rabi Hamiltonian is related to the
semiclassical one, Eq.~(\ref{dodon.H}). The infinite-dimensional state space of
the cavity electromagnetic field is spanned by the Fock states $|n\rangle$%
, where $n=0,1,2,\ldots $, so the state of the total atom-field system can
be expanded as%
\begin{equation}
|\Psi (t)\rangle =\sum_{n=0}^{\infty }\left[ A_{n}(t)|g\rangle
+B_{n}(t)|e\rangle \right] \otimes |n\rangle \,.  \label{dodon.state}
\end{equation}%
Defining a new wavefunction $|\Psi _{1}(t)\rangle $ as $|\Psi (t)\rangle
=\exp \left( -i\omega \hat{n}t\right) |\Psi _{1}(t)\rangle $, we trivially
find that $|\Psi _{1}(t)\rangle $ obeys the Schr\"{o}dinger equation with
the time-dependent Hamiltonian%
\begin{equation}
\hat{H}_{QRM}^{(1)}/\hbar =\frac{\Omega }{2}\hat{\sigma}_{z}+g\left( \hat{a}%
e^{-i\omega t}+\hat{a}^{\dagger }e^{i\omega t}\right) \left( \hat{\sigma}%
_{+}+\hat{\sigma}_{-}\right) \,.
\end{equation}%
Let us assume that the cavity is prepared initially in the coherent state
(without loss of generality, we assume that $\alpha $ is real)
\begin{equation}
|\alpha \rangle =e^{-\alpha ^{2}/2}\sum_{n=0}^{\infty }\frac{\alpha ^{n}}{%
\sqrt{n!}}|n\rangle \,.
\end{equation}%
Recalling that $\hat{a}|\alpha \rangle =\alpha |\alpha \rangle $, let us
postulate that \emph{the cavity field remains permanently in this state},
no matter what. Then, the total system density operator is given by $\hat{%
\rho}_{tot}=\hat{\rho}\otimes |\alpha \rangle \langle \alpha |$, where $\hat{%
\rho}$ is the qubit's density operator, and the qubit dynamics is governed
by the Hamiltonian%
\begin{equation}
\hat{H}_{q}/\hbar =\mathrm{Tr}_{f}\left[ \hat{H}_{QRM}^{(1)}/\hbar \otimes
|\alpha \rangle \langle \alpha |\right] =\frac{\Omega }{2}\hat{\sigma}%
_{z}+g\alpha \left( e^{-i\omega t}+e^{i\omega t}\right) \left( \hat{\sigma}%
_{+}+\hat{\sigma}_{-}\right) \,,
\end{equation}%
which reduces to the semiclassical Hamiltonian~(\ref{dodon.H}) with the
identification $G=2g\alpha $. In the above formula $\mathrm{Tr}_{f}$ denotes
the partial trace over the field. For a rigorous analysis of the transition
from QRM to SRM see Refs.~\cite{dodon.shus1,dodon.shus2}.

For the numeric simulations, we make use of the Gaussian profile of the
photon number distribution of $|\alpha \rangle $ for large average photon
numbers, $\left\langle \hat{n}\right\rangle =\left\langle \alpha |\hat{n}%
|\alpha \right\rangle =\alpha ^{2}\gg 1$. From the asymptotic expansion%
\begin{equation}
\left\vert \langle n|\alpha \rangle \right\vert ^{2}\simeq \frac{1}{\sqrt{%
2\pi }\Delta n}\exp \left[ -\left( \frac{n-\left( \left\langle \hat{n}%
\right\rangle -1/2\right) }{\sqrt{2}\Delta n}\right) ^{2}\right] \,,
\end{equation}%
where $\Delta n=\sqrt{\left\langle n^{2}\right\rangle -\left\langle
n\right\rangle ^{2}}=\alpha $ is the standard deviation, we see that the
photon number probability is centered at $\left\langle \hat{n}\right\rangle
-1/2$ and goes to zero when $\left\vert n-\alpha ^{2}\right\vert \gg \alpha $%
. In this work we consider the dynamical regime in which there is no
creation nor annihilation of a significant number of excitations (a tiny
number of such excitations is naturally generated due to the
counter-rotating terms of the Rabi Hamiltonian, but they are negligible
unless $g\alpha \sim \omega $ or there are resonant external modulations
\cite{dodon.ro1,dodon.ro2,dodon.ro3}), so we truncate the state~(\ref{dodon.state}) as%
\begin{equation}
|\Psi (t)\rangle =\sum_{n=N_{1}}^{N_{2}}\left( A_{n}(t)|g\rangle
+B_{n}(t)|e\rangle \right) \otimes |n\rangle \,,
\end{equation}%
where $N_{1}$ and $N_{2}$ are chosen so that for the initial state $|\Psi
(0)\rangle =|g\rangle \otimes |\alpha \rangle $ one has $A_{n}(0)<10^{-10}$
for $n\notin \left[ N_{1},N_{2}\right] $. For concreteness, the values we
use are (K stands for $10^{3}$):%
\begin{equation*}
\begin{tabular}{|c|c|c|}
\hline
Value of $\left\langle \hat{n}\right\rangle =\alpha ^{2}$ & $N_{1}$ & $N_{2}$
\\ \hline
5K & $4,309$ & $5,723$ \\ \hline
10K & $9,019$ & $11,013$ \\ \hline
20K & $18,746$ & $21,280$ \\ \hline
30K & $28,299$ & $31,733$ \\ \hline
40K & $38,036$ & $41,996$ \\ \hline
\end{tabular}%
\end{equation*}%
In the following section we solve the coupled ordinary differential
equations for $A_{n}$ and $B_{n}$, obtained from the Schr\"{o}dinger
equation with the Hamiltonian $\hat{H}_{QRM}$, via the Runge-Kutta-Verner
fifth-order and sixth-order method. We consider the initial state $|g\rangle
\otimes |\alpha \rangle $, exact atom--field resonance ($\Delta =0$) and set
the one-photon coupling constant through the condition $g\alpha /\omega
=10^{-2}\pi $, which corresponds to the constant $\pi $-pulse duration $%
\omega T_{\pi }=50$ for any value of $\alpha \gg 1$.

\subsection{Evolution of $P_{e}$ in QRM}

\begin{figure}[tbh]
\begin{center}
\includegraphics[width=0.99\textwidth]{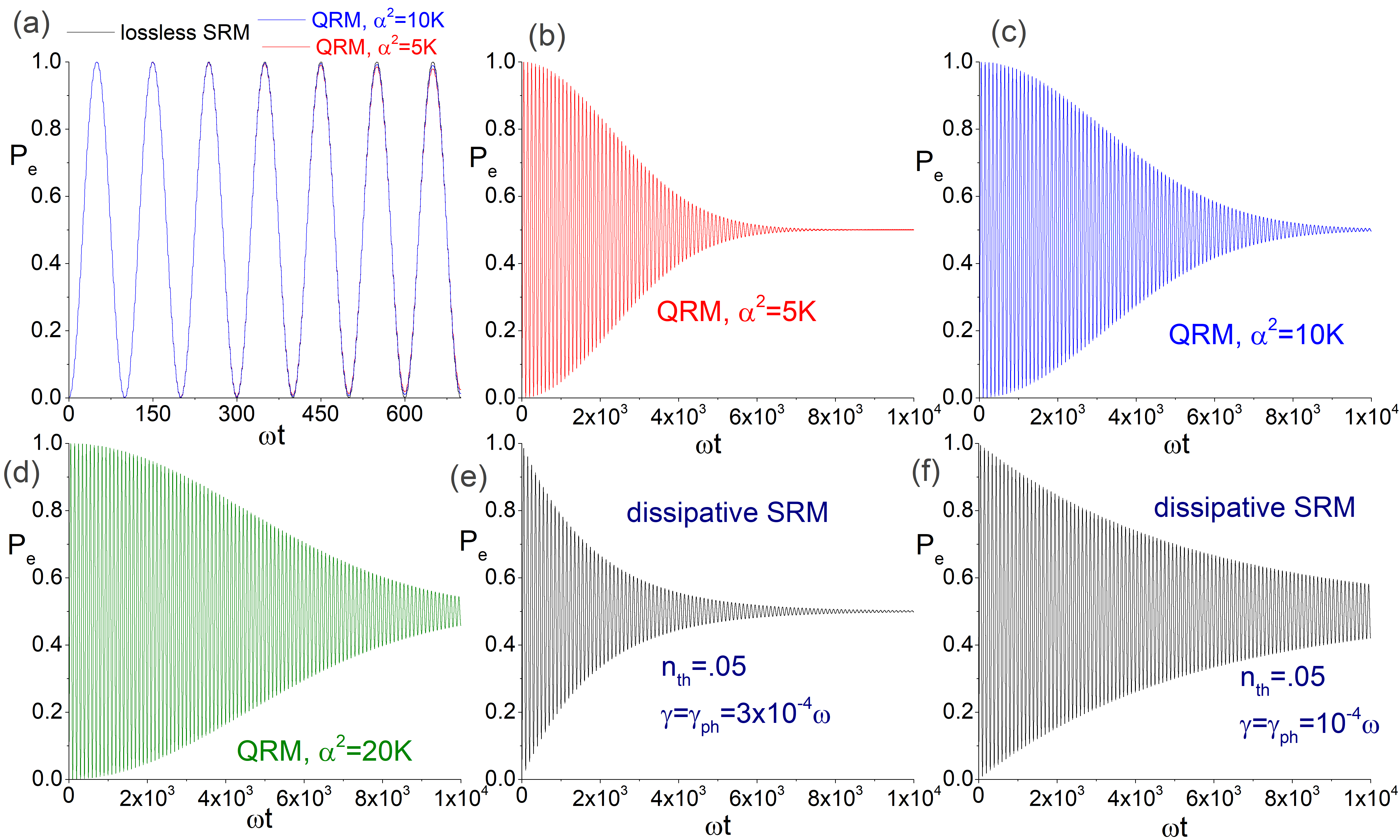} {}
\end{center}
\caption{{}Atomic excitation probability as function of time in different
scenarios. a) Comparison of the lossless SRM (black line, representing the
exact numeric solution and the semi-analytic one, which are
indistinguishable in this case) to the QRM for $\protect\alpha ^{2}=5$K and $%
10$K. b-d) QRM for $\protect\alpha ^{2}=5$K, $10$K and $20$K. e-f) Numeric
solution of the SRM in the presence of dissipation for the dissipative
parameters indicated in the plots. Notice that the behavior of $P_{e}$ is
quantitatively different in the dissipative SRM and the lossless QRM,
although in both cases $P_{e}$ undergoes a collapse.}
\label{dodon.fig2}
\end{figure}

In Fig.~\ref{dodon.fig2} we show the dynamics of the atomic excitation probability
for SRM and QRM, obtained by solving numerically the master equation (\ref%
{dodon.eme}) and the Schr\"{o}dinger equation with the Hamiltonian~(\ref{dodon.QRM}),
respectively. Panel~\ref{dodon.fig2}a shows the dynamics for initial times without
dissipation. We see that for $t\lesssim 5T_{\pi }$ the semiclassical results
are almost indistinguishable from the quantum ones, even for a relatively
weak coherent state with $\alpha ^{2}=5$K (and the agreement becomes better
as $\alpha $ increases). Therefore, for a few Rabi oscillations and
\textquotedblleft classical\textquotedblright\ fields (with $\alpha
^{2}\gtrsim 5$K, for our parameters) the semi-analytic formula can be used
to estimate (with a negligible error) the behaviour of $P_{e}$.

\begin{figure}[tbh]
\begin{center}
\includegraphics[width=0.99\textwidth]{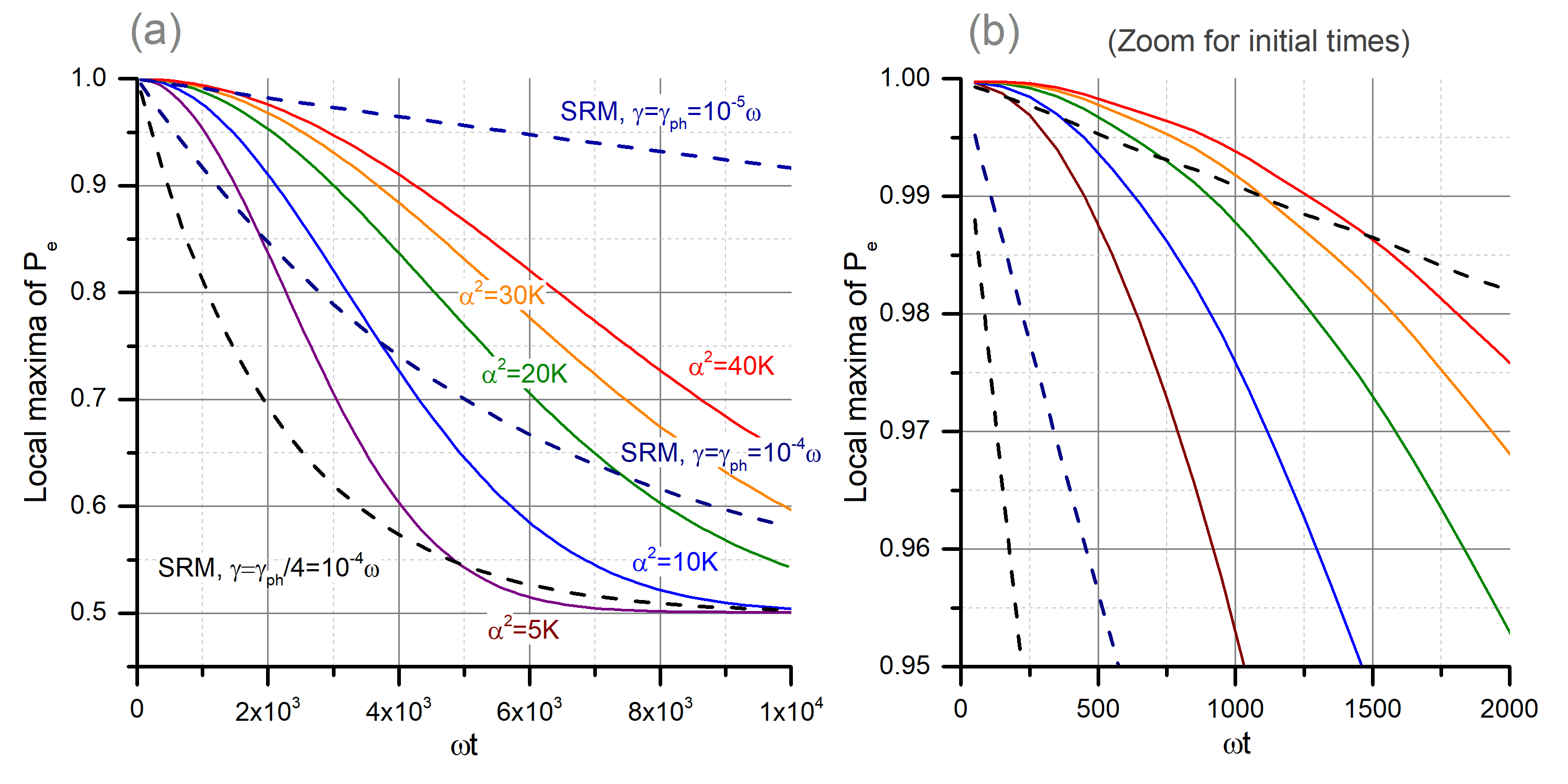} {}
\end{center}
\caption{{}Behaviour of the local maxima of $P_{e}$ as function of time for
dissipative SRM (dashed lines, with $n_{th}=0.05$) and lossless QRM (solid
lines). The dissipative parameters and the values of $\protect\alpha ^{2}$
are indicated in the figure. The panel (b) is the zoom of (a) for initial
times. Notice that in the QRM the collapse time $T_{\func{col}}$ scales as $%
\protect\alpha $.}
\label{dodon.fig3}
\end{figure}

However, for larger times the predictions of quantum and semiclassical
models start to differ qualitatively due to the characteristic collapse
behaviour, present in the quantum case at the typical timescales $\omega T_{%
\func{col}}\sim \pi \omega /2g=\alpha \omega T_{\pi }$, which for our
parameters means $\omega T_{\func{col}}\sim 50\alpha $. This estimate can be
obtained following the standard arguments of destructive interference of the
quantum Rabi oscillations corresponding to different photon numbers \cite%
{dodon.scully,dodon.eberly}, and is attested by the Figures~\ref{dodon.fig2} and~\ref{dodon.fig3}.
By the same arguments, the revival time is estimated as
\begin{equation}
\left( \omega T_{rev}\right) \sim \frac{2\pi \alpha \omega }{g}=4\alpha
^{2}(\omega T_{\pi })=200\alpha ^{2}\,,
\end{equation}%
so even for $\alpha ^{2}=5$K it would occur at $\omega T_{rev}\sim 10^{6}$.
For such long time intervals the dissipation would alter completely the
dynamics, so it seems unrealistic to study theoretically the revivals for
such coherent states without taking the dissipation into account, hence we
do not pursue this subject here.

Notice also that the collapse behaviour of $P_{e}$ in the quantum model is
qualitatively different from the semiclassical collapse due to dissipation.
This is illustrated in panels~\ref{dodon.fig2}e and~\ref{dodon.fig2}f, where we solved
numerically the master equation for the parameters indicated in the figure.
Although the atomic population also goes asymptotically to 1/2, the behaviour
of the envelope (denoting the local maxima of $P_{e}$ during the Rabi
oscillations) is clearly different. This is most clearly seen in Figure \ref%
{dodon.fig3}, where we illustrate the behaviour of the local maxima of $P_{e}$.
This figure confirms that the collapse time indeed scales approximately as $%
\omega T_{\func{col}}\sim 50\alpha $ (for instance, notice that the time for
which $P_{e}=0.6$ for $\alpha ^{2}=20$K is precisely twice the corresponding
time for $\alpha ^{2}=5$K), and the decay of the envelope is strikingly
different in the QRM and dissipative SRM.

\subsection{Backreaction on the cavity field and entanglement}

Finally, we proceed to study some important features overlooked in the
semiclassical model: the backreaction of the atom on the cavity field
(which modifies the field state) and the atom--field entanglement (which
lowers the qubit purity). We notice that the backreaction of the atom on the field has been studied in the past \cite{dodon.ashhab}, however, we believe that the numerical analysis
presented here complements nicely the previous results.

\begin{figure}[tbh]
\begin{center}
\includegraphics[width=0.99\textwidth]{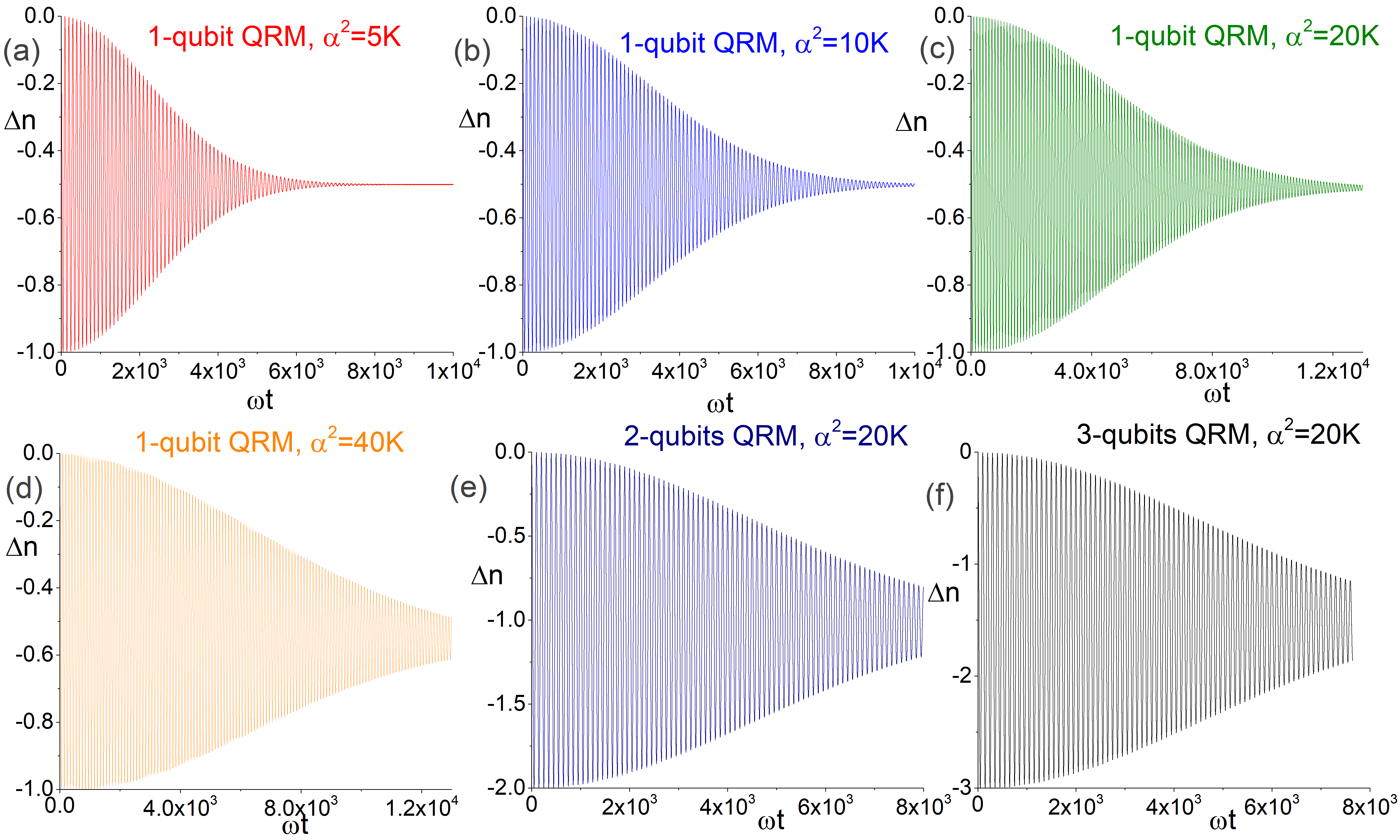} {}
\end{center}
\caption{Time behavior of the variation of the cavity field average photon
number in the lossless QRM. a-d) 1-qubit QRM with $\protect\alpha ^{2}=5$K, $%
10$K, $20$K and $30$K. e-f) QRM with two and three identical noninteracting
qubits, respectively, and $\protect\alpha ^{2}=20$K. Notice that in all the
cases $\Delta n$ exhibits a collapse behavior.}
\label{dodon.fig4}
\end{figure}

In Fig.~\ref{dodon.fig4} we show the variation of the cavity average photon
number, $\Delta n=\left\langle \hat{n}(t)\right\rangle -\left\langle \hat{n}%
(0)\right\rangle $ as a function of time. In the panels~\ref{dodon.fig4}a -- \ref%
{dodon.fig4}d we consider a single-qubit QRM with $\alpha ^{2}=5$K, $10$K, $20$K
and $30$K, respectively. As expected intuitively, the cavity field exchanges
one photon with the qubit, and the average photon number also undergoes a
collapse process on the same timescale as the qubit. In the panels~\ref{dodon.fig4}%
e and~\ref{dodon.fig4}f we consider the scenario with two and three noninteracting
qubits (with identical parameters) that interact with the cavity field in
the initial coherent state with $\alpha ^{2}=20$K. We verified that, in this
case, the dynamics of each qubit is almost indistinguishable from the
dynamics in the 1-qubit setup (shown in Fig.~\ref{dodon.fig2}c). The cavity field
exchanges two and three photons with the atoms, respectively, while also
undergoing a collapse process.

\begin{figure}[tbh]
\begin{center}
\includegraphics[width=0.99\textwidth]{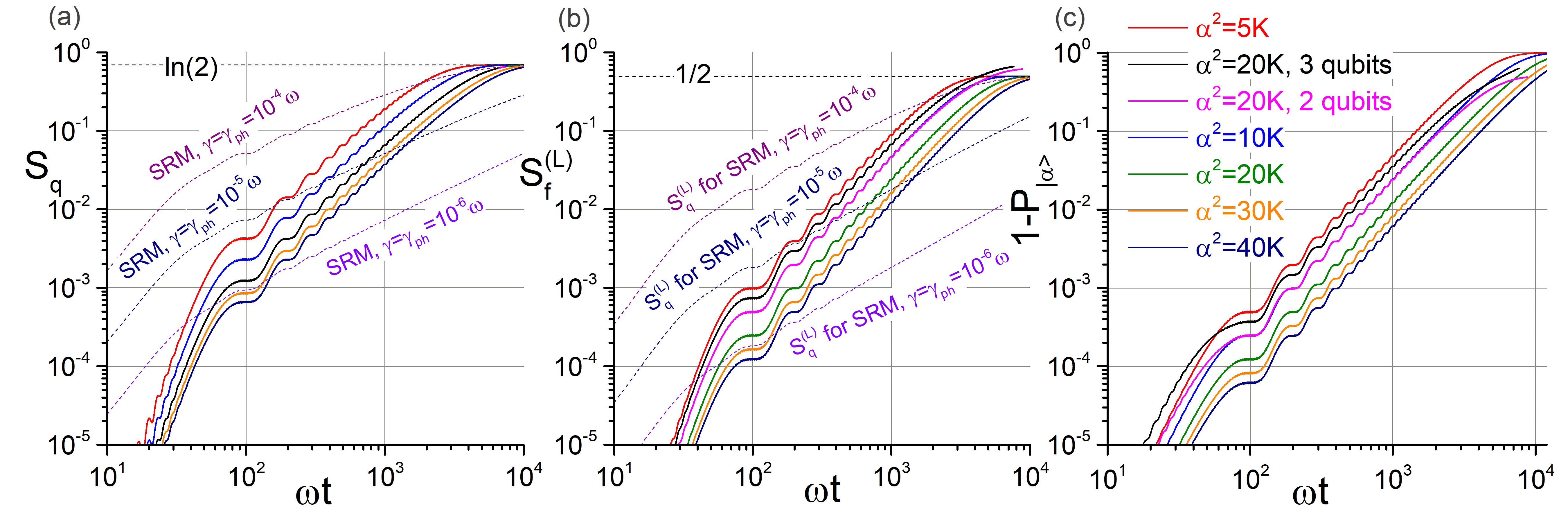} {}
\end{center}
\caption{{}a) Dynamics of the von Neumann entropy of a single qubit in the
dissipative SRM (dashed lines, assuming $n_{th}=0.05$) and lossless QRM
(solid lines). The parameters and the values of $\protect\alpha ^{2}$ are
indicated in the figure. b) Linear entropy of the cavity field state. c)
Probability that the cavity field state does not remain in the original
coherent state $|\protect\alpha (t)\rangle $, where $\protect\alpha (t)=%
\protect\alpha e^{-i\protect\omega t}$. Note: some curves are incomplete, because the simulations were still running at the moment of the writing of this chapter.}
\label{dodon.fig5}
\end{figure}

It is obvious that the qubit(s) become entangled with the cavity field
during the evolution, but without numeric calculations it is impossible to
infer the amount of entanglement and its dynamical behaviour. To characterize
the entanglement of a \emph{pure bipartite system} it is enough to study the
von Neumann or the linear entropies of any subsystem, $S_{i}\equiv -\limfunc{%
Tr}\left( \hat{\rho}_{i}\ln \hat{\rho}_{i}\right) $ and $S_{i}^{(L)}\equiv 1-%
\limfunc{Tr}\left( \hat{\rho}_{i}^{2}\right) $, respectively, where $\hat{%
\rho}_{i}$ is the reduced density operator of any subsystem (this is a
straightforward consequence of the Schmidt decomposition~\cite{dodon.orszag,dodon.nielsen}). For a multipartite system, on the other hand, these
entropies only measure the entanglement between the subsystem $i$ and the
rest of the system. We chose to study the von Neumann entropy for the qubit,
$S_{q}$, and the linear entropy for the cavity field, $S_{f}^{(L)}$. Notice
that for the single-qubit QRM we have $S_{q}^{(L)}=S_{f}^{(L)}$, so in this
case $S_{f}^{(L)}$ also quantifies the purity of the atom, which is an
important figure of merit in Quantum Information protocols. In the panels %
\ref{dodon.fig5}a and~\ref{dodon.fig5}b, the solid lines show $S_{q}$ and $S_{f}^{(L)}$
as a function of time for QRM (in the case of 2- and 3-qubits QRM, these are
the entropies of any qubit and the field, respectively), while the dashed
lines show the von Neumann and the linear entropies of the qubit in the
dissipative SRM. We see that the qubit's purity loss due to the exchange of
photons with the quantized cavity field is as important as the dissipative
effects in SRM when $\gamma ,\gamma _{ph}\lesssim 10^{-6}\omega $ (as
expected, for larger $\alpha $ the loss of purity is smaller at a given
time). We also see that the atom(s) and the field become entangled from the
very beginning, and do not disentangle even after a complete Rabi
oscillation, attaining a high degree of entanglement for the times of the
order of $T_{\func{col}}$ (with the maximum allowed values of the entropies,
$S_{q,\max }=\ln 2$ and $S_{f,\max }^{(L)}=1/2$, in the case of a single
qubit). Finally, these plots also show that the purity of the atom, due to
the quantized nature of the field, differs from 1 by roughly $10^{-4}$ at
the moment of the first complete Rabi oscillation (for the chosen
parameters).

\begin{figure}[tbh]
\begin{center}
\includegraphics[width=0.99\textwidth]{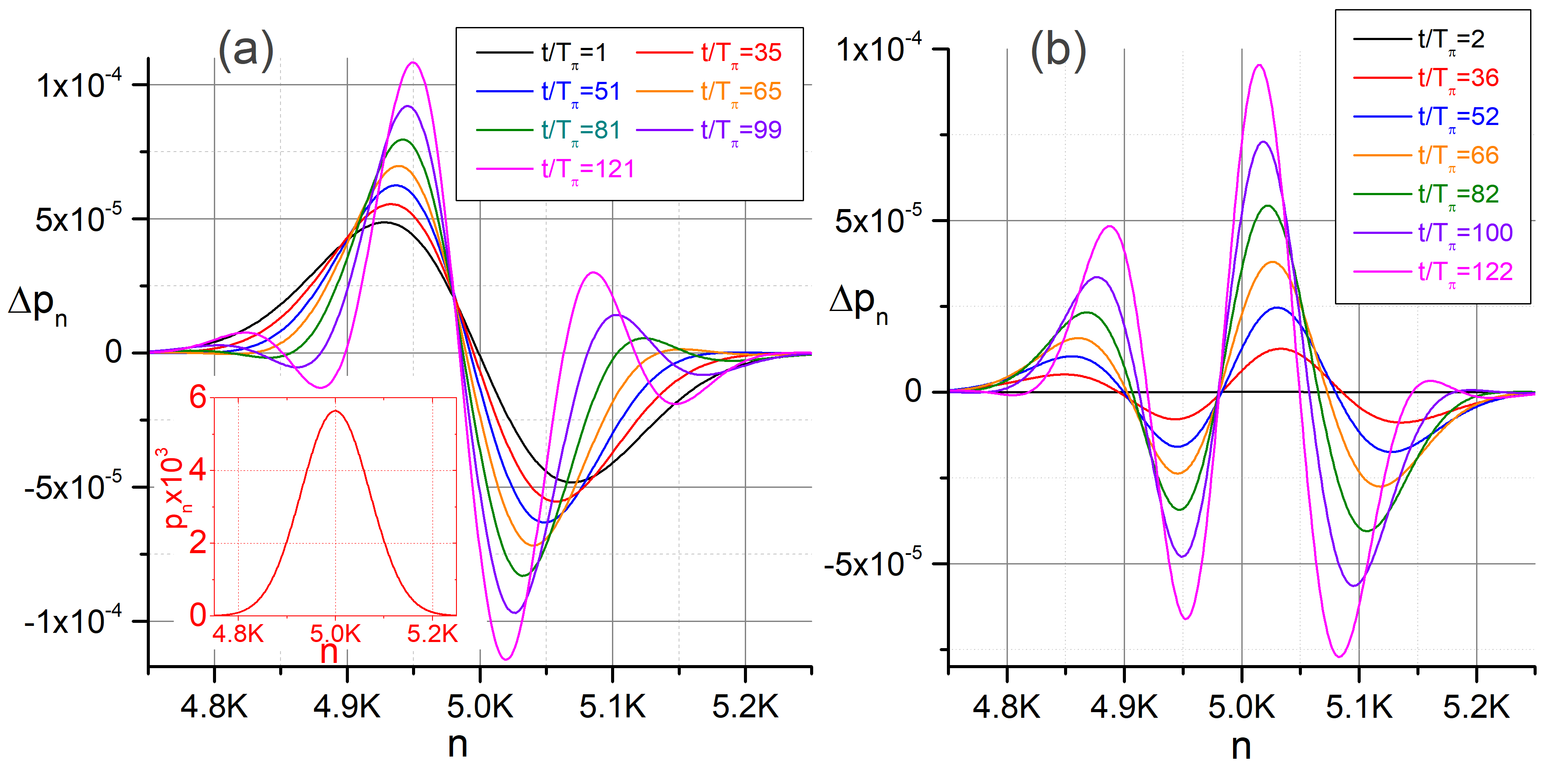} {}
\end{center}
\caption{a) {}Modification of the photon number distribution of the cavity
state (compared to the initial one) for $\protect\alpha ^{2}=5$K and odd
multiples of $T_{\protect\pi }$, when the qubit excitation probability
attains local maxima. The inset shows the original photon number
distribution. b) Modification of the photon number statistics at even
multiples of $T_{\protect\pi }$, when the atomic excitation probability
attains local minima. }
\label{dodon.fig6}
\end{figure}

We already noticed that the average photon number decreases during the
atom-field interaction, so an interesting question is what is the
probability of the cavity field remaining in the coherent state $|\alpha
(t)\rangle $, where $\alpha (t)=\alpha e^{-i\omega t}$ is the field
amplitude (in the interaction picture we are adopting here) that would hold
in the absence of the atom(s). To answer this question, we calculated
numerically the probability $P_{|\alpha \rangle }=\limfunc{Tr}\left[ \hat{%
\rho}|\alpha e^{-i\omega t}\rangle \langle \alpha e^{-i\omega t}|\right] $
of the field remaining in the state $|\alpha (t)\rangle $, and plotted the
quantity $1-P_{|\alpha \rangle }$ in the panel~\ref{dodon.fig5}c. This panel shows
that the cavity field state is drastically altered during the evolution and
does not return to the freely evolving state $|\alpha (t)\rangle $ for the
considered time intervals, as opposed to the SRM assumption that the field
remains in the same state throughout the evolution.

\begin{figure}[tbh]
\begin{center}
\includegraphics[width=0.99\textwidth]{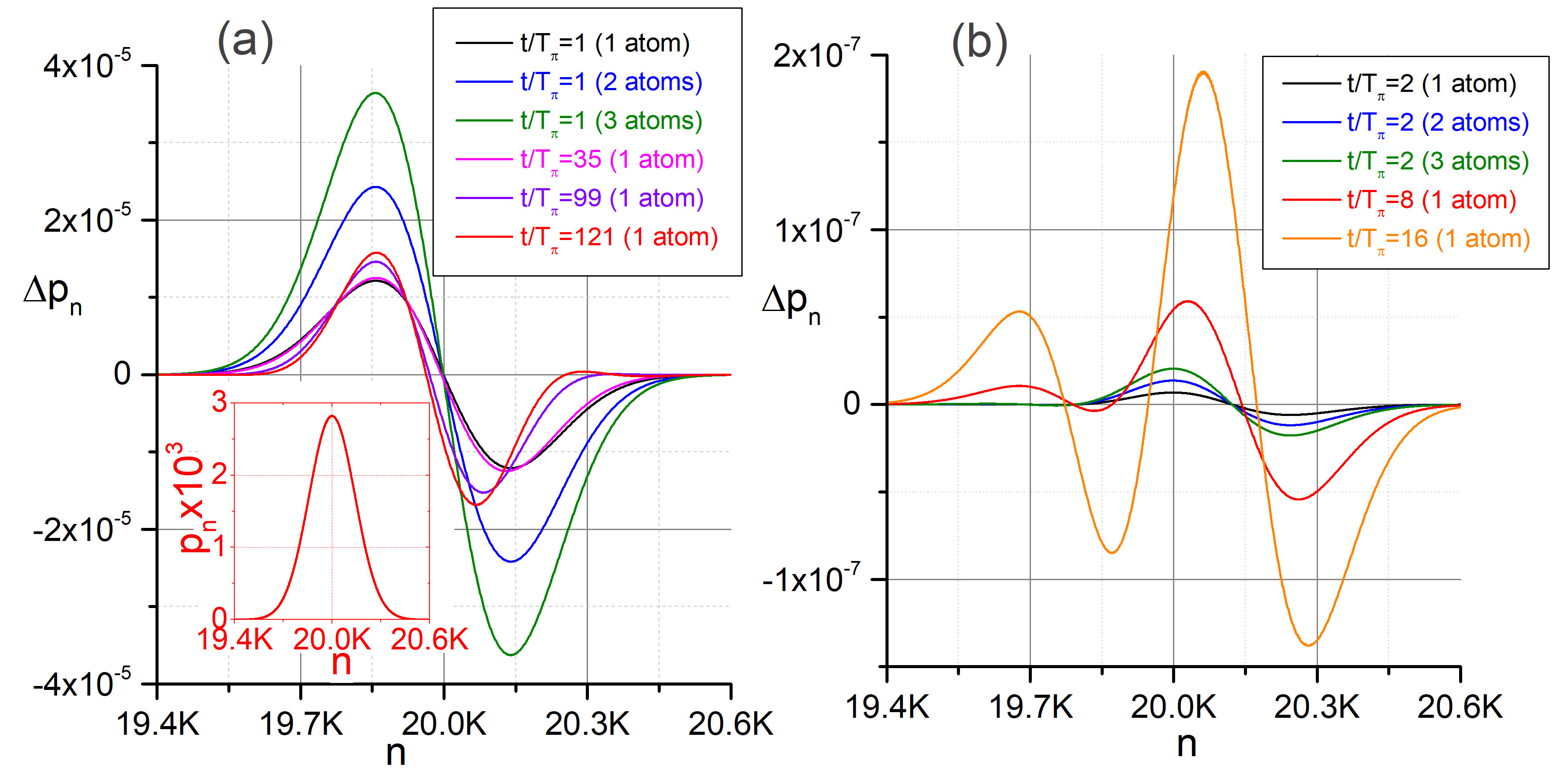} {}
\end{center}
\caption{{}Similar to Fig. \protect\ref{dodon.fig6} but for $\protect\alpha %
^{2}=20 $K and contemplating the scenarios with 1, 2 and 3 identical
noninteracting qubits coupled to the cavity field.}
\label{dodon.fig7}
\end{figure}

\begin{figure}[tbh]
\begin{center}
\includegraphics[width=0.99\textwidth]{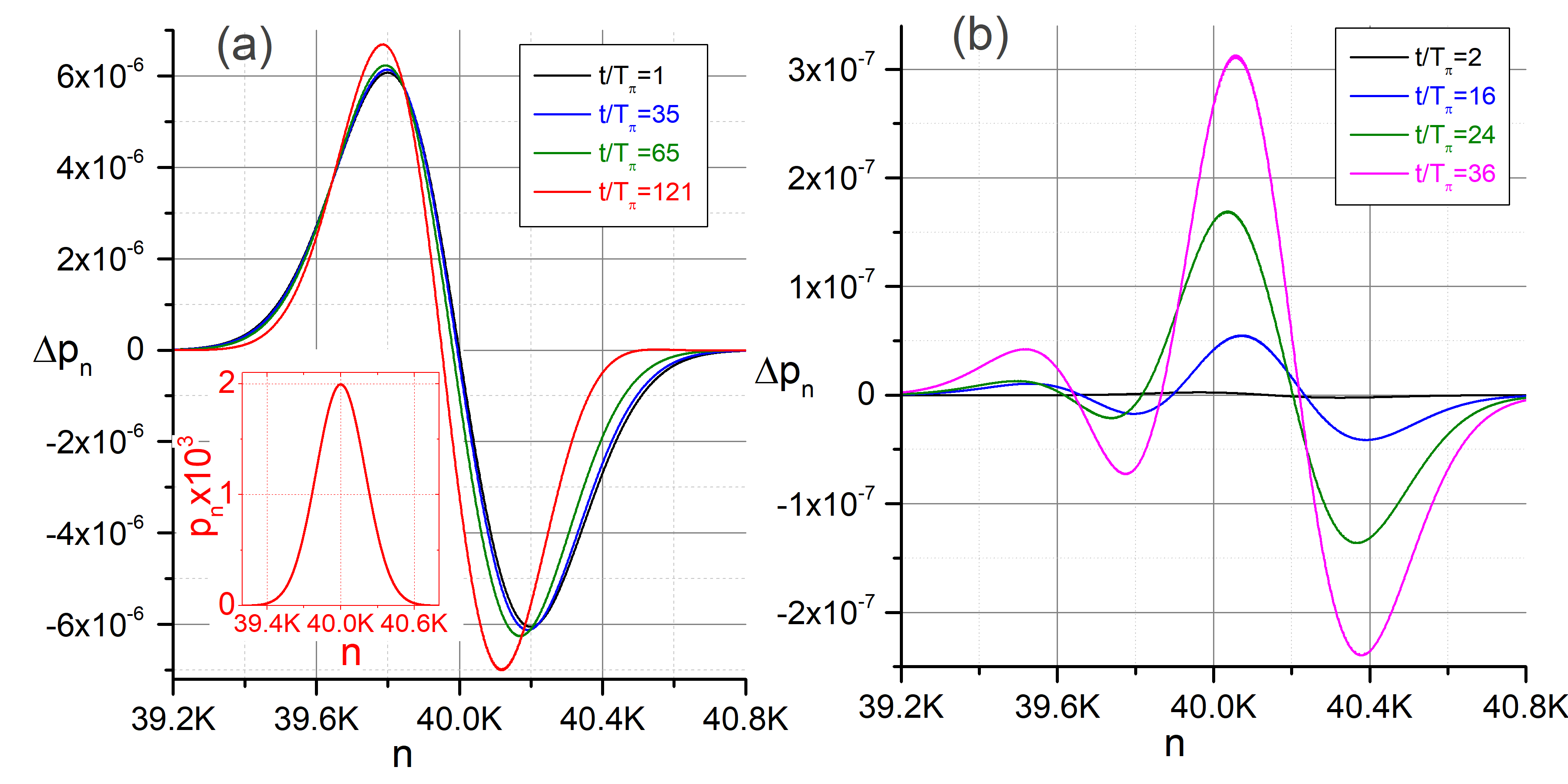} {}
\end{center}
\caption{{}Similar to Fig. \protect\ref{dodon.fig6} but for $\protect\alpha %
^{2}=40 $K.}
\label{dodon.fig8}
\end{figure}

To understand why $P_{|\alpha \rangle }\rightarrow 0$ for large times, we
studied the photon number distribution for different times. In Fig. \ref%
{dodon.fig6} we show the results for $\alpha ^{2}=5$K. Panel~\ref{dodon.fig6}a shows $%
\Delta p_{n}=p_{n}(t)-p_{n}(0)$, which is the difference between the photon
number distribution at the time $t$, $p_{n}(t)=\limfunc{Tr}\left( \hat{\rho}%
(t)|n\rangle \langle n|\right) $, and the initial photon number distribution
(shown in the inset), where $t$ is some odd multiple of $T_{\pi }$
(corresponding to the local maxima of the atomic excitation probability).
The panel~\ref{dodon.fig6}b shows $\Delta p_{n}$ for some even multiples of $%
T_{\pi }$, when the atomic excitation probability attains a local minimum.
In Figures~\ref{dodon.fig7} and~\ref{dodon.fig8} we do a similar analysis for $\alpha
^{2}=20$K (considering the 2- and 3-qubit QRM, as well) and $\alpha ^{2}=40$%
K, respectively.

We see that for initial times, $t\lesssim 10T_{\pi }$, the behaviour of $%
\Delta p_{n}$ is intuitive. At the moments of a maximum of $P_{e}$ (odd
multiples of $T_{\pi }$), $\Delta p_{n}$ is positive for $n<\alpha ^{2}$ and
negative for $n>\alpha ^{2}$; for even multiples of $T_{\pi }$, when the
qubit attains the minimum excitation probability, $\Delta p_{n}$ is nearly
zero. However, for larger times the behaviour becomes more complex, and for
even multiples of $T_{\pi }$ the field state does not return to its initial
statistics. This is consistent with the emergence of the collapse behaviour
of $P_{e}$ and $\Delta n$ for large times, since the field state does not
return to its original state after a complete Rabi oscillation. We also note
that, for a given time $t$, $\Delta p_{n}(t)$ becomes smaller when $\alpha $
increases (recalling that we assume that $g\alpha $ is constant), so for $%
\alpha \rightarrow \infty $ and initial times ($t\lll T_{\func{col}}$) the
tiny modifications of $\Delta p_{n}$ become undetectable experimentally and
the semiclassical assumption that the cavity field is not altered during the
evolution becomes justified. However, our analysis shows that as the time
increases, the semiclassical approximation progressively loses its validity due to a significant modification of the photon number statistics and the
entanglement between the cavity field and the atom(s).

\subsection{Dissipation in the quantum regime}

\begin{figure}[tbh]
\begin{center}
\includegraphics[width=0.99\textwidth]{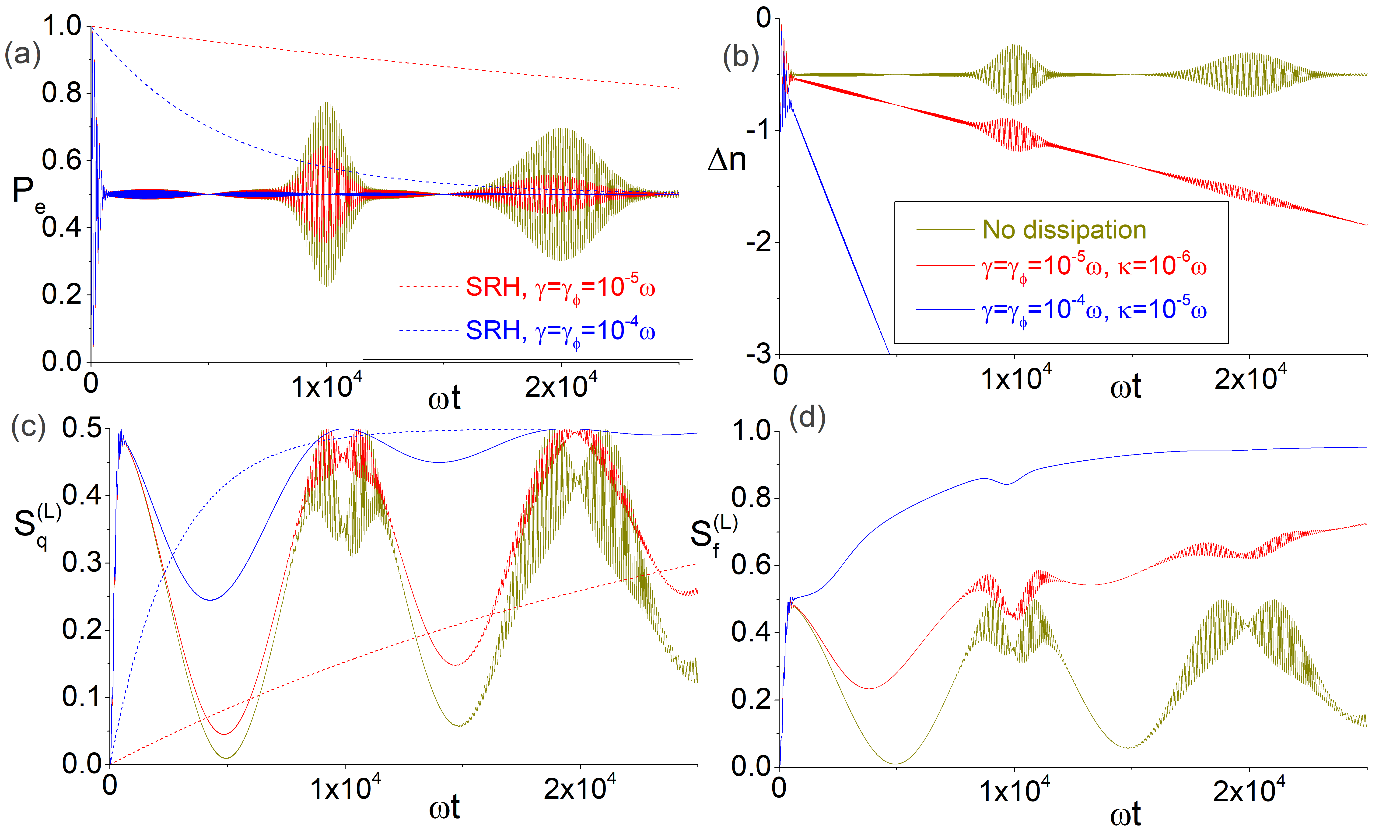} {}
\end{center}
\caption{Dissipative quantum and semiclassical Rabi models. Solid lines
illustrate the time evolution of $P_{e}${} (panel a), $\Delta n$ (panel b), $%
S_{q}^{(L)}$ (panel c) and $S_{f}^{(L)}$ (panel d) according to the QRM for
different dissipative parameters in the master equation (\protect\ref{dodon.fun}),
assuming $\protect\alpha ^{2}=50$, $\protect\omega T_{\protect\pi }=50$ and $%
n_{c}=n_{th}=0.05$. The dashed lines illustrate the dissipative behaviour of
the envelope of $P_{e}$ (panel a) and $S_{q}^{(L)}$ (panel c) according to
the SRM.}
\label{dodon.Fig9}
\end{figure}

Before concluding this work, let us study numerically how the dissipation
affects the system dynamics in the quantum regime, when the total density
operator obeys the Lindblad master equation
\begin{eqnarray}
\frac{\partial \hat{\rho}}{\partial t} &=&-i\left[ \hat{H}_{QRM}/\hbar ,\hat{%
\rho}\right] +\frac{\gamma }{2}(n_{th}+1)\mathcal{D}[\hat{\sigma}_{-}]\hat{%
\rho}+\frac{\gamma }{2}n_{th}\mathcal{D}[\hat{\sigma}_{+}]\hat{\rho}+\frac{%
\gamma _{\phi }}{2}\mathcal{D}[\hat{\sigma}_{z}]\hat{\rho}  \notag \\
&&+\frac{\kappa }{2}(n_{c}+1)\mathcal{D}[\hat{a}]\hat{\rho}+\frac{\kappa }{2}%
n_{c}\mathcal{D}[\hat{a}^{\dagger }]\hat{\rho}\,,  \label{dodon.fun}
\end{eqnarray}%
where $\kappa $ is the cavity damping rate and $n_{c}=\left[ e^{\hbar \omega
/k_{B}T}-1\right] ^{-1}$ is the average thermal photon number for the cavity
frequency $\omega $ and the temperature $T$. Since for the density matrix
the number of elements grows quadratically with the number of photons, we
managed to solve numerically the master equation~(\ref{dodon.fun}) only for a
relatively weak initial coherent state $|g,a\rangle $ with $\alpha ^{2}=50$
(maintaining the previous parameters $\Delta =0$ and $g\alpha /\omega
=10^{-2}\pi $). In Fig.~\ref{dodon.Fig9} we show the results for the
dissipationless case (dark yellow lines) and two different sets of the
dissipative parameters: $\gamma =\gamma _{\phi }=10^{-5}\omega $, $\kappa
=10^{-6}\omega $ (red lines) and $\gamma =\gamma _{\phi }=10^{-4}\omega $, $%
\kappa =10^{-5}\omega $ (blue lines), where we assumed $n_{th}=n_{c}=0.05$.
As expected, in the unitary case there are revivals of $P_{e}$ and $\Delta n$
for the integer multiples of $\left( \omega T_{rev}\right) \sim 200\alpha
^{2}=10^{4}$. Besides, we see that the linear entropies $S_{q}^{(L)}$ and $%
S_{f}^{(L)}$ do not saturate at some constant values as one could misinterpret
from Fig.~\ref{dodon.fig5}; instead, they exhibit oscillatory behaviours on the
timescales of the order of $T_{rev}$, going nearly to zero halfway between
the revivals. Such behaviours were not captured by Fig.~\ref{dodon.fig5} because
the necessary time intervals would be $\omega t\gtrsim 100\alpha ^{2}$, and
for such long times the effects of dissipation would be dominant. Typical
behaviours of the QRM in the presence of dissipation are illustrated by the
red and blue solid lines in Fig.~\ref{dodon.Fig9}; for the considered dissipative
parameters, the dynamics loses the resemblance with the unitary case for
times $\omega t\gtrsim 10^{4}$, and we expect a similar conclusion to hold
also for larger values of $\alpha $ (provided $g\alpha $ is maintained
constant). Our numeric solution of the master equation shows that, in the
presence of dissipation, the revivals of $P_{e}$ and $\Delta n$, as well as
the oscillatory behavior of the linear entropies, are first attenuated and
then completely eliminated for large times.

Fig.~\ref{dodon.Fig9} also shows that, for the assumed dissipative parameters,
the collapse of $P_{e}$ in the QRM occurs much faster than the semiclassical
collapse in the SRM due to the dissipation. The envelope of $P_{e}$ and the
linear entropy $S_{q}^{(L)}$ (obtained by solving numerically the
dissipative SRM with the same parameters as the dissipative QRM) are
depicted by the dashed lines in the panels~\ref{dodon.Fig9}a and~\ref{dodon.Fig9}c,
respectively. We see that even in the dissipative case the SRM utterly fails
in describing the qubit behaviour when it interacts with a relatively weak
coherent state, but for such small values of $\alpha ^{2}\sim 50$ the
failure of SRM was completely expected.

\section{Summary}

In this work we revisited the semiclassical and quantum Rabi models at the
exact atom-field resonance. We illustrated how the textbook sinusoidal
formula for the Rabi oscillations loses its validity as the duration of the $%
\pi $-pulse decreases below $T_{\pi }\sim 100\omega ^{-1}$ and presented
simple analytic and semi-analytic expressions that describe more accurately
the behaviour of the atomic excitation probability. Moreover, we carried
out detailed numeric simulations of the quantum Rabi model for the initial
coherent state with a large average photon number, ranging from $5.000$ to $%
40.000$, showing that for initial times the quantum dynamics approaches the
semiclassical one (in what concerns the atomic Rabi oscillations), while for
larger times the atomic excitation probability inevitably exhibits a
collapse behaviour (which is, of course, absent in the semiclassical model).
The behaviour of such collapses is qualitatively different from the
oscillatory decay of the atomic excited state due to the Markovian damping
and dephasing.

We believe that the main contribution of this work to the vast literature on
the quantum and semiclassical Rabi models was to study numerically the
backreaction of the atom on the state of the cavity field, as well as
quantifying the atom--field entanglement in the context of the QRM. We
showed that the excitation of one or more noninteracting qubits coupled to a
single field mode is accompanied by the extraction of a single photon, per
atom, from the cavity field; it was also shown that such exchange of
excitations ceases for large times due to the collapses of $P_{e}$ and $%
\Delta n$. Over the course of time the atom becomes entangled with the
field, and the system evolves to a maximally entangled state upon the
collapse of $P_{e}$ and $\Delta n$ (as attested by the maximum values of von
Neumann and linear entropies, studied in Figure~\ref{dodon.fig5}); however, from
our simulations carried for small values of $\alpha $, we expect that for
large times (of the order of the revival times), the atom--field
entanglement should exhibit an oscillatory behaviour. Finally, we studied how
the photon number statistics is modified during the evolution, showing
that after a complete Rabi oscillation the cavity field does not return to
its original state, and for times of the order of the collapse time the
probability of finding the field in the original (freely evolving) coherent
state tends to zero (although it could exhibit eventual revivals for very
long times).

\section*{Acknowledgment}

A. P. C. acknowledges the financial support by the Brazilian agency Coordena%
\c{c}\~{a}o de Aperfei\c{c}oamento de Pessoal de N\'{\i}vel Superior (CAPES,
Finance Code~001). A. D. acknowledges partial financial support of the
Brazilian agencies CNPq (Conselho Nacional de Desenvolvimento Cient\'{\i}%
fico e Tecnol\'{o}gico) and Funda\c{c}\~{a}o de Apoio \`{a} Pesquisa do
Distrito Federal (FAPDF, grant number 00193-00001817/2023-43).

\renewcommand{\bibname}{References} \begingroup
\let\cleardoublepage\relax

\endgroup 


\begin{thebibliography}{99}
\bibitem{dodon.mi} M. Le Bellac, \emph{A Short Introduction to Quantum Information and
Quantum Computation} (Cambridge University Press, Cambridge, 2012).

\setlength{\itemsep}{0em} \setlength{\parskip}{0em}

\bibitem{dodon.mat} Y. Sch\"{o}n, J. N. Voss, M. Wildermuth, et al. Rabi
oscillations in a superconducting nanowire circuit. npj Quantum Mater.
\textbf{5}, 18 (2020).

\bibitem{dodon.calib} Y. Huang, M. T. Amawi, F. Poggiali, F. Shi, J. Du and F.
Reinhard, Calibrating single-qubit gates by a two-dimensional Rabi
oscillation. AIP Advances \textbf{13}, 035226 (2023).

\bibitem{dodon.rabi1} I. I. Rabi. On the Process of Space Quantization. Phys. Rev.
\textbf{49}, 324 (1936).

\bibitem{dodon.rabi2} I. I. Rabi. Space quantization in a gyrating magnetic field.
Phys. Rev. \textbf{51}, 652 (1937).

\bibitem{dodon.merlin} R. Merlin. Rabi oscillations, Floquet states, Fermi's
golden rule, and all that: Insights from an exactly solvable two-level
model. Am. J. Phys. \textbf{89}, 26 (2021).

\bibitem{dodon.boyd} R. W. Boyd. \emph{Nonlinear Optics} (Academic Press, London, 2nd
Edition, 2003).

\bibitem{dodon.scully} M. O. Scully and M. S. Zubairy. \emph{Quantum Optics} (Cambridge University Press, Cambridge, 1997).

\bibitem{dodon.shore} B. W. Shore. Coherent manipulation of atoms using laser
light. Acta Physica Slovaca \textbf{58}, 243 (2008).

\bibitem{dodon.graham} R. Graham and M. H\"{o}hnerbach. Two-state system coupled
to a boson mode: Quantum dynamics and classical approximations. Z. Physik B
- Cond. Matt. \textbf{57}, 233 (1984).

\bibitem{dodon.munz} M. Munz and G. Marowsky. Rabi-oscillations without
rotating-wave approximation. Z. Phys. B - Condensed Matter \textbf{63}, 131
(1986).

\bibitem{dodon.Liu} J. Liu and Z.-Y. Li. Interaction of a two-level atom with
single-mode optical field beyond the rotating wave approximation. Opt. Expr.
\textbf{22}, 28671 (2014).

\bibitem{dodon.Lu} L.-Z. Lu, D.-Q. Wen, S.-J. Jiang and X.-Y. Yu.
Interaction between the ultrashort pulse and two-level medium beyond the
rotating wave approximation. Eur. Phys. J. D \textbf{70}, 184 (2016).

\bibitem{dodon.ashhab} S. Ashhab. Landau-Zener-Stueckelberg interferometry with driving fields in the quantum regime. J. Phys. A: Math. Theor. \textbf{50}, 134002 (2017).

\bibitem{dodon.castanos} L. O. Casta\~{n}os. Simple, analytic solutions of the
semiclassical Rabi model. Opt. Commun. 430, 176 (2019).

\bibitem{dodon.ad3} I. Sainz, A. Garc\'{\i}a and A. B. Klimov. Effective and
efficient resonant transitions in periodically modulated quantum systems.
Quantum Reports \textbf{3}, 1 (2021).

\bibitem{dodon.ad7} I. Sainz, A. B. Klimov and C. Saavedra. Effective Hamiltonian
approach to periodically perturbed quantum optical systems. Phys. Lett. A
\textbf{351}, 26 (2006).

\bibitem{dodon.marinho1} A. Marinho and A. Dodonov. Analytic approach for
dissipative semiclassical Rabi model under parametric modulation, in A.
Dodonov and C. C. H. Ribeiro (Eds.), Proceedings of the Second International
Workshop on Quantum Nonstationary Systems, pp. 195--210 (LF Editorial, S\~{a}%
o Paulo, 2024). ISBN: 978-65-5563-446-4.

\bibitem{dodon.marinho3} A. Marinho and A. V. Dodonov. Approximate analytic
solution of the dissipative semiclassical Rabi model under parametric
multi-tone modulations. Phys. Scr. \textbf{99}, 125117 (2024).

\bibitem{dodon.rev} Q.~Xie, H.~Zhong, M.~T~Batchelor and C.~Lee, The quantum Rabi
model: solution and dynamics. J. Phys. A.: Math. Theor. \textbf{50}, 113001
(2017).

\bibitem{dodon.solano} D. Braak, Q.-H. Chen, M. T. Batchelor and
E. Solano. Semi-classical and quantum Rabi models: in celebration of 80
years. J. Phys. A: Math. and Theor. \textbf{49}, 300301 (2016).

\bibitem{dodon.braak} D. Braak. Integrability of the Rabi model. Phys. Rev. Lett.
\textbf{107}, 100401 (2011).

\bibitem{dodon.rson} J. Larson and Th. K. Mavrogordatos. \emph{The Jaynes-Cummings Model and Its Descendants} (IOP Publishing, Bristol, 2021). https://arxiv.org/abs/2202.00330 (2024).

\bibitem{dodon.shus1} E. K. Irish and A. D. Armour. Defining the semiclassical
limit of the quantum Rabi Hamiltonian. Phys. Rev. Lett. \textbf{129}, 183603
(2022).

\bibitem{dodon.shus2} H. F. A. Coleman and E. K. Twyeffort. Spectral and dynamical
validity of the rotating-wave approximation in the quantum and semiclassical
Rabi models. J. Opt. Soc. Am. B 41, C188 (2024).

\bibitem{dodon.marinho2} A. Marinho, M. V. S. de Paula and A. V. Dodonov.
Approximate analytic solution of the dissipative semiclassical Rabi model
near the three-photon resonance and comparison with the quantum behaviour.
Phys. Lett. A \textbf{513}, 129608 (2024).

\bibitem{dodon.ad8} H. Carmichael. \emph{An Open System Approach to Quantum Optics}
(Springer, Berlin, 1993).

\bibitem{dodon.vogel} W. Vogel and D.-G. Welsch. \emph{Quantum Optics} (Wiley, Berlin,
2006).

\bibitem{dodon.orszag} M. Orszag. \emph{Quantum Optics} (Springer, Berlin, 2nd Edition,
2008).

\bibitem{dodon.ro1} A. V. Dodonov. Errors in zero-excitation state preparation due
to anti-rotating terms in two-atom Markovian cavity QED. Phys. Scr. \textbf{%
82}, 055401 (2010).

\bibitem{dodon.ro2} A. V. Dodonov. Mean excitation numbers due to the
anti-rotating term in cavity QED under Lindbladian dephasing. Phys. Scr.
\textbf{86}, 025405 (2012).

\bibitem{dodon.ro3} I. M. de Sousa and A. V. Dodonov. Microscopic toy model for
the cavity dynamical Casimir effect. J. Phys. A: Math. Theor. \textbf{48},
245302 (2015).

\bibitem{dodon.eberly} J. H. Eberly, N. B. Narozhny and J. J. Sanchez-Mondragon. Periodic spontaneous collapse and revival in a simple quantum model. Phys. Rev. Lett. \textbf{44}, 1323 (1980).

\bibitem{dodon.nielsen} M. A. Nielsen and I. L. Chuang. \emph{Quantum Computation and
Quantum Information} (Cambridge University Press, Cambridge, 10th Edition, 2010).
\end{thebibliography}
\end{document}